\batchmode

\documentclass[12pt,epsf]{article}

\usepackage{times}
\usepackage{graphicx}
\usepackage{amsfonts}
\usepackage{amsmath}
\usepackage{amssymb}
\usepackage{amstext}
\usepackage{amsthm}

\usepackage{epsfig}
\usepackage{color}
\psfigdriver{dvips}
\usepackage{latexsym}
\usepackage[matrix,curve]{xypic}
\usepackage{rotating}
\rotdriver{dvips}

\begin{document}

\baselineskip 6mm
\renewcommand{\thefootnote}{\fnsymbol{footnote}}


\newcommand{\nc}{\newcommand}
\newcommand{\rnc}{\renewcommand}


\rnc{\baselinestretch}{1.24}    
\setlength{\jot}{6pt}       
\rnc{\arraystretch}{1.24}   

\makeatletter
\rnc{\theequation}{\thesection.\arabic{equation}}
\@addtoreset{equation}{section}
\makeatother



\nc{\be}{\begin{equation}}

\nc{\ee}{\end{equation}}

\nc{\bea}{\begin{eqnarray}}

\nc{\eea}{\end{eqnarray}}

\nc{\xx}{\nonumber\\}

\nc{\ct}{\cite}

\nc{\la}{\label}

\nc{\eq}[1]{(\ref{#1})}

\nc{\newcaption}[1]{\centerline{\parbox{6in}{\caption{#1}}}}

\nc{\fig}[3]{

\begin{figure}
\centerline{\epsfxsize=#1\epsfbox{#2.eps}}
\newcaption{#3. \label{#2}}
\end{figure}
}


\def\CA{{\cal A}}
\def\CC{{\cal C}}
\def\CD{{\cal D}}
\def\CE{{\cal E}}
\def\CF{{\cal F}}
\def\CG{{\cal G}}
\def\CH{{\cal H}}
\def\CK{{\cal K}}
\def\CL{{\cal L}}
\def\CM{{\cal M}}
\def\CN{{\cal N}}
\def\CO{{\cal O}}
\def\CP{{\cal P}}
\def\CR{{\cal R}}
\def\CS{{\cal S}}
\def\CU{{\cal U}}
\def\CV{{\cal V}}
\def\CW{{\cal W}}
\def\CY{{\cal Y}}
\def\CZ{{\cal Z}}


\def\IB{{\hbox{{\rm I}\kern-.2em\hbox{\rm B}}}}
\def\IC{\,\,{\hbox{{\rm I}\kern-.50em\hbox{\bf C}}}}
\def\ID{{\hbox{{\rm I}\kern-.2em\hbox{\rm D}}}}
\def\IF{{\hbox{{\rm I}\kern-.2em\hbox{\rm F}}}}
\def\IH{{\hbox{{\rm I}\kern-.2em\hbox{\rm H}}}}
\def\IN{{\hbox{{\rm I}\kern-.2em\hbox{\rm N}}}}
\def\IP{{\hbox{{\rm I}\kern-.2em\hbox{\rm P}}}}
\def\IR{{\hbox{{\rm I}\kern-.2em\hbox{\rm R}}}}
\def\IZ{{\hbox{{\rm Z}\kern-.4em\hbox{\rm Z}}}}


\def\a{\alpha}
\def\b{\beta}
\def\d{\delta}
\def\ep{\epsilon}
\def\ga{\gamma}
\def\k{\kappa}
\def\l{\lambda}
\def\s{\sigma}
\def\t{\theta}
\def\w{\omega}
\def\G{\Gamma}


\def\half{\frac{1}{2}}
\def\dint#1#2{\int\limits_{#1}^{#2}}
\def\goto{\rightarrow}
\def\para{\parallel}
\def\brac#1{\langle #1 \rangle}
\def\curl{\nabla\times}
\def\div{\nabla\cdot}
\def\p{\partial}


\def\Tr{{\rm Tr}\,}
\def\det{{\rm det}}


\def\vare{\varepsilon}
\def\zbar{\bar{z}}
\def\wbar{\bar{w}}
\def\what#1{\widehat{#1}}


\def\ad{\dot{a}}
\def\bd{\dot{b}}
\def\cd{\dot{c}}
\def\dd{\dot{d}}
\def\so{SO(4)}
\def\bfr{{\bf R}}
\def\bfc{{\bf C}}
\def\bfz{{\bf Z}}

\begin{titlepage}


\hfill\parbox{3.7cm} {{\tt arXiv:1810.12291}}

\vspace{15mm}

\begin{center}
{\Large \bf  Expanding Universe and Dynamical Compactification \\ Using Yang-Mills Instantons}

\vspace{10mm}

Kyung Kiu Kim ${}^{a}$\footnote{kimkyungkiu@sejong.ac.kr},
Seoktae Koh ${}^{b}$\footnote{kundol.koh@jejunu.ac.kr}
and Hyun Seok Yang ${}^c$\footnote{hsyang@sogang.ac.kr}
\\[10mm]

${}^a$ {\sl Department of Physics and Astronomy, Sejong University, Seoul 05006, Korea}

${}^b$ {\sl Department of Science Education, Jeju National University, Jeju, 63243, Korea}

${}^c$ {\sl Center for Quantum Spacetime, Sogang University, Seoul 121-741, Korea}

\end{center}

\thispagestyle{empty}

\vskip1cm


\centerline{\bf ABSTRACT}
\vskip 4mm
\noindent

We consider an eight-dimensional Einstein-Yang-Mills theory to explore
whether Yang-Mills instantons formed in extra dimensions
can induce the dynamical instability of our four-dimensional spacetime.
We show that the Yang-Mills instantons in extra dimensions can trigger
the expansion of our universe in four-dimensional spacetime
as well as the dynamical compactification of extra dimensions.
We also discuss a possibility to realize a reheating mechanism
via the quantum back-reaction from the contracting tiny internal space
with a smeared instanton.
\\


Keywords: Cosmic inflation, Yang-Mills instanton, Dynamical compactification

\vspace{1cm}

\today

\end{titlepage}

\renewcommand{\thefootnote}{\arabic{footnote}}
\setcounter{footnote}{0}

\section{Introduction}

The idea of inflation postulates a period of accelerated expansion during the Universe's earliest stages
in order to explain the initial conditions for the hot big bang model \cite{ceu-inflation}.
It has subsequently been given a much more important role as the currently-favored candidate
for the origin of structure in the Universe,
such as galaxies, galaxy clusters and cosmic microwave background anisotropies \cite{cden-pert}.
The recent CMB observations have unambiguously proven
the theory of the quantum origin of the universe structure \cite{inf-test}.
See also Chapters 8 \& 9 in \cite{kolb-turner}.
Quantum fluctuations in the metric itself during inflation leads to an almost scale-invariant
spectrum of primordial gravitational waves as well as density perturbations, which may
contribute to the low multipoles of the CMB anisotropy.
Hence the goal of inflationary cosmology is to understand the origin and early evolution of the universe
and explain the observed large-scale structure with predictions for future observations.
However, there are important unsolved problems in inflationary cosmology.
For example, a few are listed as the fluctuation problem, the super-Planck-scale physics problem,
the singularity problem and the cosmological constant problem \cite{Brandenberger:2007qi}.

The inflation scenario so far has been formulated in the context of effective field
theory coupled to general relativity.
The simplest way to implement an inflationary phase is to assume some scalar field
which carries significant vacuum energy.
However, the expected outcome of inflation can easily change if we vary the initial conditions,
change the shape of the inflationary potential, and often leads to eternal inflation
and a multiverse \cite{Ijjas:2013vea}.
In order for inflation to really answer the questions about our universe
(for example, nature of the ``inflaton" field, initial conditions and slow-roll
conditions for inflation, graceful exit and reheating),
we must identify the microphysical origin of this scalar field and its inflation potential.
How is it connected to fundamental physics and
what determines the details of the action including the gravity, kinetic, and potential
sectors? In order to solve such problems, it is necessary to derive the theory of inflation
from a more sophisticated fundamental theory.

The leading candidate for a fundamental theory of quantum gravity
is string theory. In order for both string theory and inflation to be correct, one should
demonstrate explicitly that string theory offers a correct cosmological description of
the universe and incorporates inflation \cite{string-cos,Baumann:2014nda}.
Most of the effort in string theory to realize our universe has been devoted to constructing
de Sitter vacua \cite{Kachru:2003aw,Kachru:2003sx}.
But it turns out \cite{Hertzberg:2007wc,Brennan:2017rbf} that it is very difficult to obtain a meta-stable
de Sitter vacuum and it is fair to say that these scenarios have not yet been rigorously shown
to be realized in string theory. Moreover, the landscape of string theory gives a vast range
of choices for how our universe may fit in a consistent quantum theory of gravity.
Given these difficulties in obtaining de Sitter vacua in string theory,
it has been recently conjectured \cite{swampland} that string theory does not admit de Sitter critical points.
It has been also claimed \cite{Agrawal:2018own} that inflationary models are generically in tension with
the Swampland criteria which states the conditions that a low-energy effective theory has
to satisfy to have any consistent UV completion inside a theory of quantum gravity.
However, in these attempts to construct de Sitter-like vacua,
a low-energy effective theory has been largely based on scalar fields
such as vacuum moduli arising in string theory, which introduce
notorious problems such as the moduli stabilization \cite{Kachru:2003sx,Balasubramanian:2005zx}.

In this paper, we suggest an alternative inflationary model using Yang-Mills instantons in
a higher-dimensional gauge theory to set the physics of inflation on a more solid basis.
We embed the cosmic inflation into the general framework of a well-defined field theory
that describes particles and their interactions without introducing any scalar field
as well as an {\it ad hoc} inflation potential.
A basic idea is easy to understand. Consider a $(4+1)$-dimensional
$SU(2)$ Yang-Mills gauge theory. If one considers a static solution in this theory with
the Hamilton gauge $A_0 = 0$,
the Hamiltonian $\mathcal{H}= - \frac{1}{4 g_{YM}^2} \int d^4 y \mathrm{Tr} F_{\alpha\beta}F^{\alpha\beta}$
is equal to the four-dimensional Euclidean action,
so $\mathcal{H}$ is minimized with Yang-Mills instantons.
This means that Yang-Mills instantons give rise to a non-zero (quantized) energy
$\mathcal{H}= \frac{8 \pi^2 n}{g_{YM}^2}$ in $(4+1)$-dimensional spacetime where
a positive integer $n$ is the instanton number.
We can generalize this argument to the eight-dimensional Yang-Mills gauge theory
so that Yang-Mills instantons are formed in four-dimensional internal space but homogeneous in our four-dimensional spacetime.
Then the instanton action corresponds to a (quantized) constant energy density
$\rho= \frac{8 \pi^2 n}{g_{YM}^2}$ in our four-dimensional spacetime. If the gauge
theory is coupled to the eight-dimensional gravity, such instantons
act as a (quantized) cosmological constant in four-dimensional Universe.
Therefore it is interesting to ask whether Yang-Mills instantons formed in
extra dimensions can trigger a cosmic inflation in our four-dimensional spacetime.
We scrutinize the dynamical instability of our four-dimensional spacetime induced by Yang-Mills instantons.
Our analysis indicates that the Yang-Mills instantons in internal space
can act as an anti-gravitating matter source in our four-dimensional Universe.
An accelerating expansion from and a dynamical compactification of extra dimensions are an old idea explored
in many literatures \cite{inf-exd,inf-exd2,dyn-comp}
(see also Chapter 11.4 in \cite{kolb-turner})
and similar ideas using monopoles and instantons in extra dimensions have appeared in \cite{old-refs,sim-idea}.

This paper is organized as follows. In section 2, we explain our idea on how Yang-Mills instantons formed
in a four-dimensional internal space generate a (quantized) cosmological constant
in four-dimensional spacetime in an approximation of ignoring their gravitational back-reaction.
In section 3, we couple the eight-dimensional Yang-Mills gauge theory to Einstein gravity to incorporate
the gravitational back-reaction from Yang-Mills instantons in extra dimensions.
We derive underlying equations describing dynamical evolution of the eight-dimensional spacetime
triggered by Yang-Mills instantons. Thus our setup is essentially different from \cite{sim-idea} in which
the eight-dimensional spacetime is static.
In section 4, we numerically solve the evolution equations.
We find that the dynamical behavior of the internal space and four-dimensional
spacetime is opposite so that the internal space is contracting if our four-dimensional spacetime
is expanding and vice versa \cite{inf-exd2,dyn-comp}. We point out an interesting picture that a dynamical compactification
of extra dimensions may be realized through the cosmic inflation of our four-dimensional spacetime
and the quantum back-reaction from the tiny internal space may be used to realize
a reheating mechanism and a hot big bang at early universe.
The internal space may have had a more direct role in the cosmological
evolution of our universe  \cite{inf-exd}.
In section 5, we discuss another interesting features of our model and generalizations
to a ten-dimensional spacetime applicable to string theory.

\section{Yang-Mills instantons and quantized cosmological constant}

Consider an eight-dimensional spacetime $\mathcal{M}_8$ whose metric is given by
\begin{equation}\label{8-metric}
    ds^2 = G_{MN} dX^M dX^N = e^A \otimes e^A,
\end{equation}
where $X^M = (x^\mu, y^\alpha), \; M, N = 0, 1, \cdots, 7; \mu,\nu = 0,1,2,3;
\alpha, \beta =4,5,6,7$, are local coordinates on $\mathcal{M}_8$ and $e^A = (e^m, e^a)
\; A, B = 0, 1, \cdots, 7; m,n = 0,1,2,3; a, b =4,5,6,7$, are orthonormal vielbeins in
$\Gamma(T^* \mathcal{M}_8)$. Let $\pi: E \to \mathcal{M}_8$ be a $G$-bundle
over $\mathcal{M}_8$ whose curvature is given by
\begin{eqnarray} \label{g-curvature}
  F &=& dA + A \wedge A = \frac{1}{2} F_{MN} (X) dX^M \wedge dX^N \\
    &=& \frac{1}{2} \Big( \partial_M A_N - \partial_N A_M + [A_M, A_N]  \Big) dX^M \wedge dX^N
\end{eqnarray}
where $A = A^i_M (X) \tau^i dX^M = \big( A_\mu (x,y) dx^\mu, A_\alpha (x,y) dy^\alpha \big)$
is a connection one-form of the $G$-bundle $E$ and
$\tau^i \;\big(i=1, \cdots, \mathrm{dim}(G) \big)$ are Lie algebra generators
obeying the commutation relation
\begin{equation}\label{lie-comm}
    [\tau^i, \tau^j] = f^{ijk} \tau^k.
\end{equation}
We choose a normalization $\mathrm{Tr} \tau^i \tau^j = - \delta^{ij}$.
The action for the eight-dimensional Yang-Mills theory on a curved manifold $\mathcal{M}_8$
is then defined by
\begin{equation}\label{8-action}
    S_{YM} = \frac{1}{4 g_{YM}^2} \int_{\mathcal{M}_8} d^8 X \sqrt{-G} G^{MP} G^{NQ} \mathrm{Tr} F_{MN} F_{PQ}.
\end{equation}

In order to see whether Yang-Mills instantons formed in extra dimensions give rise to
a vacuum energy which triggers the expansion of our Universe in four-dimensional spacetime,
let us consider a simple geometry $\mathcal{M}_8 = \mathcal{M}_{3,1} \times X_4$ with a product metric
\begin{eqnarray}\label{prod-metric}
    ds^2 &=& G_{MN} dX^M dX^N = g_{\mu\nu} (x) dx^\mu dx^\nu + h_{\alpha\beta} (y) dy^\alpha dy^\beta \xx
    &=& e^m \otimes e^m + e^a \otimes e^a.
\end{eqnarray}
For this product geometry, the action \eq{8-action} takes the form
\begin{equation}\label{8-prod-action}
    S_{YM} = \frac{1}{4 g_{YM}^2} \int_{\mathcal{M}_{3,1}} d^4 x  \sqrt{-g}
    \int_{X_4} d^4 y \sqrt{h} \mathrm{Tr} \Big( g^{\mu\nu} g^{\rho\sigma} F_{\mu\nu} F_{\rho\sigma}
    + 2 g^{\mu\nu} h^{\alpha\beta} F_{\mu\alpha} F_{\nu\beta}
    + h^{\alpha\gamma} h^{\beta\delta} F_{\alpha\beta} F_{\gamma\delta} \Big).
\end{equation}
We are interested in the gauge field configuration given by
\begin{equation}\label{8-gauge}
    A_\mu (x, y) = 0, \qquad A_\alpha (x,y) = A_\alpha (y),
\end{equation}
for which the above action reduces to
\begin{equation}\label{8-4-action}
    S_{YM} = \frac{1}{4 g_{YM}^2} \int_{\mathcal{M}_{3,1}} d^4 x  \sqrt{-g}
    \int_{X_4} d^4 y \sqrt{h} h^{\alpha\gamma} h^{\beta\delta} \mathrm{Tr} F_{\alpha\beta} F_{\gamma\delta}.
\end{equation}
The question is if there exists any gauge field configuration for which the four-dimensional action
along the internal space $X_4$ becomes a non-zero constant, i.e.,
\begin{equation}\label{4int-action}
    I_n \equiv - \int_{X_4} d^4 y \sqrt{h} h^{\alpha\gamma} h^{\beta\delta}
    \mathrm{Tr} F_{\alpha\beta} F_{\gamma\delta} = \mathrm{constant}.
\end{equation}
It is well-known \cite{rajaraman} that the four-dimensional gauge fields satisfying the condition \eq{4int-action} are
precisely Yang-Mills instantons obeying the self-duality equation
\begin{equation}\label{self-dual-eq}
  F_{\alpha\beta} = \pm \frac{1}{2} \frac{\varepsilon^{\xi\eta \gamma\delta}}{\sqrt{h}}
  h_{\alpha\xi} h_{\beta\eta} F_{\gamma\delta}.
\end{equation}
In this case, $I_n = 32 \pi^2 n$ with $n \in \mathbb{N}$ and the action \eq{8-4-action} can be written as
\begin{equation}\label{4in-action}
    S_{YM} = - \frac{1}{4 g_{YM}^2} \int_{\mathcal{M}_{3,1}} d^4 x  \sqrt{-g} I_n.
\end{equation}
Therefore we see that the Yang-Mills instantons in Eq. \eq{4int-action} generate the coupling with
the {\it quantized} cosmological constant $\Lambda$ in the four-dimensional spacetime.
Since $[g_{YM}^2] = L^4$ in eight dimensions, it may be instructive to rewrite \eq{4in-action} as
\begin{equation}\label{cc-action}
    S_{\Lambda} = - \frac{1}{8 \pi G_4} \int_{\mathcal{M}_{3,1}} d^4 x  \sqrt{-g} \Lambda
\end{equation}
where $G_4$ is the four-dimensional Newton constant and
\begin{equation}\label{cc}
    \Lambda \equiv \frac{2 \pi G_4}{g_{YM}^2} I_n \geq 0
\end{equation}
has the correct dimension of the cosmological constant in four dimensions, i.e.,
$[\Lambda] = L_H^{-2}$.

In sum, if Yang-Mills instantons are formed in $X_4$, their instanton number
behaves like a (quantized) cosmological constant in $\mathcal{M}_{3,1}$ in an approximation
ignoring the gravitational back-reaction due to Yang-Mills instantons.
Hence it will be interesting to examine whether the Yang-Mills instantons in the internal space
can trigger a cosmic inflation in our four-dimensional spacetime.
In next section we will examine this idea.

\section{Dynamical spacetime from Yang-Mills instantons}

In order to investigate whether Yang-Mills instantons in the internal space $X_4$ can trigger
the cosmic inflation in the four-dimensional spacetime $\mathcal{M}_{3,1}$, and to see how
the four-dimensional internal space behaves,
let us consider the eight-dimensional Yang-Mills theory \eq{8-action} coupled to Einstein gravity.
It is described by the Einstein-Yang-Mills theory with the total action
\begin{equation}\label{total-action}
    S = \frac{1}{16 \pi G_8} \int_{\mathcal{M}_8} d^8 X \sqrt{-G} R + S_{YM}
\end{equation}
where $G_8$ is the eight-dimensional gravitational constant.
The gravitational field equations read as
\begin{equation}\label{einstein-eq}
    R_{MN} - \frac{1}{2} G_{MN} R = 8 \pi G_8 T_{MN}
\end{equation}
with the energy-momentum tensor given by
\begin{equation}\label{em-tensor}
    T_{MN} = - \frac{1}{g_{YM}^2} \mathrm{Tr} \Big( G^{PQ} F_{MP} F_{NQ}
    - \frac{1}{4} G_{MN} F_{PQ}F^{PQ} \Big).
\end{equation}
The action \eq{8-action} leads to the equations of motion for Yang-Mills gauge fields
\begin{equation}\label{ym-eom}
    G^{MN} D_M F_{NP} = 0,
\end{equation}
where the covariant derivative is defined with respect to both the Yang-Mills and gravitational
connections, i.e.,
\begin{equation}\label{cov-der}
D_M F_{NP} = \partial_M F_{NP} - {\Gamma_{MN}}^Q F_{QP} - {\Gamma_{MP}}^Q F_{NQ} + [A_M, F_{NP}]
\end{equation}
and ${\Gamma_{MN}}^P$ is the Levi-Civita connection. The equations of motion \eq{ym-eom} may be written
as a succinct form
\begin{equation}\label{cov-der2}
\frac{1}{\sqrt{-G}}\partial_M \big(\sqrt{-G}F^{MN}\big) + [A_M, F^{MN}] = 0.
\end{equation}
In the end, we need to show that the cosmic inflation triggered by the Yang-Mills instantons satisfies
both \eq{einstein-eq} and \eq{ym-eom}. Note that we have not introduced a bare cosmological constant
in contrast to \cite{sim-idea} since we do not want a static spacetime.

In order to solve the equations of motion, let us consider the following ansatz for an eight-dimensional metric
\begin{equation}\label{warp-metric}
    ds^2 = G_{MN} dX^M dX^N = g_{\mu\nu} (x) dx^\mu dx^\nu
    + e^{2 f(x)} h_{\alpha\beta} (y) dy^\alpha dy^\beta.
\end{equation}
Although we are considering a warped product metric \eq{warp-metric},
the separation such as Eq. \eq{8-prod-action} is still valid and
the action for the gauge field configuration \eq{8-gauge} reduces to
\begin{equation}\label{8-4-gaction}
    S_{YM} = \frac{1}{4 g_{YM}^2} \int_{\mathcal{M}_{3,1}} d^4 x
     \sqrt{-g} \int_{X_4} d^4 y \sqrt{h} h^{\alpha\gamma} h^{\beta\delta}
     \mathrm{Tr} F_{\alpha\beta} F_{\gamma\delta},
\end{equation}
where we used the fact that the action \eq{4int-action} is invariant under the Weyl transformation
$ h_{\alpha\beta} \to e^{2f(x)} h_{\alpha\beta}$.
Then one can see that the equations of motion \eq{ym-eom} take the simple form
\begin{equation}\label{ym-geom}
 h^{\alpha\beta} D_\alpha F_{\beta\gamma} = 0.
\end{equation}
It is easy to show that Eq. \eq{ym-geom} is automatically satisfied as far as the gauge fields obey
the self-duality equation \eq{self-dual-eq}.
In consequence, the Yang-Mills instantons satisfy the equations of motion \eq{ym-eom} even
in a warped spacetime with the metric \eq{warp-metric}.

The energy-momentum tensor \eq{em-tensor} is determined by the Yang-Mills instantons and
one finds that
\begin{eqnarray}\label{emtensor-1}
&& T_{\mu\nu} = \frac{1}{4 g_{YM}^2} \widetilde{g}_{\mu\nu} \mathrm{Tr} F_{\alpha\beta} F^{\alpha\beta}, \xx
&& T_{\alpha\beta} = - \frac{e^{-2f(x)}}{g_{YM}^2} \mathrm{Tr}
\Big( h^{\gamma\delta} F_{\alpha\gamma} F_{\beta\delta}
    - \frac{1}{4} h_{\alpha\beta} F_{\gamma\delta}F^{\gamma\delta} \Big) = 0, \\
&&  T_{\mu\alpha} = 0, \nonumber
\end{eqnarray}
where $\widetilde{g}_{\mu\nu} (x) = e^{-4f(x)} g_{\mu\nu} (x)$ and all indices are raised and lowered
with the product metric \eq{prod-metric}.
We used the fact \cite{rajaraman} that the energy-momentum tensor $ T_{\alpha\beta}$ identically vanishes for an instanton solution
satisfying Eq. \eq{self-dual-eq}. Let us denote the energy-momentum tensor $T_{\mu\nu}$ as the form
\begin{equation}\label{intanton-emtensor}
T_{\mu\nu} = - \frac{1}{g_{YM}^2} \widetilde{g}_{\mu\nu} (x) \rho_n (y),
\end{equation}
where $\rho_n (y) = - \frac{1}{4} \mathrm{Tr} F_{\alpha\beta} F^{\alpha\beta}$ is the instanton density in $X_4$
which is uniform along the four-dimensional spacetime $\mathcal{M}_{3,1}$.
After incorporating a gravitational back-reaction of instantons, the energy-momentum tensor induced by Yang-Mills instantons
does not precisely correspond to a cosmological constant in the warped product metric \eq{warp-metric}  contrary to the product metric \eq{prod-metric}.
The coupling in Eq. \eq{intanton-emtensor} implies that the behavior of four-dimensional spacetime ${\cal M}_{3,1}$ sourced by the energy-momentum tensor $ T_{\mu\nu}$
will depend on the size of internal space $X_4$ characterized by the warp factor $e^{2f(x)}$.
To be specific, for a fixed instanton density, if the internal space becomes small, i.e.,
$f(x)$ decreasing, then the energy-momentum tensor $ T_{\mu\nu}$ becomes large, i.e.,
${\cal M}_{3,1}$ more expanding, or vice versa.
This fact may be used to realize a dynamical compactification of the internal space, as we will discuss later.
Note that, for a single Yang-Mills instanton on $\mathbb{R}^4$ with the gauge group $G = SU(2)$,
the instanton density is given by \cite{rajaraman}
\begin{equation}\label{inst-den}
    \rho_1 (y) = \frac{48 \zeta^4}{[(y-y_0)^2 + \zeta^2]^4},
\end{equation}
where $y_0^\mu$ are position moduli of an instanton and $\zeta$ is a modulus of instanton size.

Under the ansatz \eq{warp-metric}, the gravitational field equations \eq{einstein-eq} read as
\begin{eqnarray}\label{geom-31}
     && R_{\mu\nu} - \frac{1}{2} g_{\mu\nu} R = 8 \pi G_8 T_{\mu\nu}, \\
     \label{geom-x}
     && R_{\mu\alpha} =0, \qquad  R_{\alpha\beta}- \frac{1}{2} e^{2f(x)} h_{\alpha\beta} R = 0.
\end{eqnarray}
For the warped product geometry \eq{warp-metric}, the Ricci tensors are given by\footnote{A useful note is
``Compendium of useful formulas" by Matthew Headrick, which can be downloaded from http://people.brandeis.edu/~headrick/}

\begin{eqnarray}\label{warp-ricci}
    && R_{\mu\nu} = R_{\mu\nu}^{(0)} - 4 (\nabla^{(g)}_\mu \partial_\nu f
    + \partial_\mu f\partial_\nu f), \xx
    && R_{\alpha\beta} = R_{\alpha\beta}^{(0)} - (\nabla^2_{(g)} f
    + 4 g^{\mu\nu} \partial_\mu f\partial_\nu f) e^{2f(x)} h_{\alpha\beta}, \\
    && R_{\mu \alpha} = 0, \nonumber
\end{eqnarray}
where $R_{\mu\nu}^{(0)}$ and $R_{\alpha\beta}^{(0)}$ are the Ricci tensors when $f=0$ and
$\nabla^{(g)}_\mu$ is a covariant derivative with respect to the metric $g_{\mu\nu}(x)$.
And the Ricci scalar is given by
\begin{equation}\label{warp-rsca}
    R = R_{(g)} + e^{-2f(x)} R_{(h)} - 8 \nabla^2_{(g)} f - 20 g^{\mu\nu} \partial_\mu f\partial_\nu f,
\end{equation}
where $R_{(g)}$ and $R_{(h)}$ are the Ricci scalars of the metrics $g_{\mu\nu}$ and $h_{\alpha\beta}$
when $f=0$, respectively. Therefore the Einstein equations take the form
\begin{eqnarray}\label{eins-11}
&& R_{\mu\nu}^{(0)} - \frac{1}{2} g_{\mu\nu} R_{(g)}
= 4 \big( \nabla^{(g)}_\mu \partial_\nu f
    + \partial_\mu f\partial_\nu f \big) - \big( 4 \nabla^2_{(g)} f
    + 10 g^{\rho\sigma} \partial_\rho f\partial_\sigma f - \frac{1}{2} e^{-2f} R_{(h)} \big) g_{\mu\nu} \xx
   && \hspace{3.3cm} - \frac{8 \pi G_8}{g_{YM}^2} e^{-4f(x)} \rho_n (y) g_{\mu\nu}, \\
\label{eins-22}
&& R_{\alpha\beta}^{(0)}  - \frac{1}{2} h_{\alpha\beta} R_{(h)} = 3  e^{2f(x)}
\Big( \frac{1}{6} R_{(g)} - \nabla^2_{(g)} f - 2 g^{\mu\nu} \partial_\mu f\partial_\nu f \Big) h_{\alpha\beta}.
\end{eqnarray}

A close inspection on the dependence of $\mathcal{M}_{3,1}$
and $X_4$ in Eq. \eq{eins-22} leads to the condition that a solution exists
only if the internal space described by the metric $h_{\alpha\beta}$
is a space with $R_{(h)}=$ constant. Then one can find a consistent solution of Eq. \eq{eins-11}
only when $\rho_1 (y) =\mathrm{constant} := \frac{48}{\zeta_c^4}$, i.e.,
\begin{equation}\label{cinst-den}
 \rho_1 (y) = - \frac{1}{4} h^{\alpha\gamma} h^{\beta\delta}
 \mathrm{Tr} F_{\alpha\beta} F_{\gamma\delta} = \frac{48}{\zeta_c^4}.
\end{equation}
We will assume that the internal space $X_4$ has a conformally flat
metric $h_{\alpha\beta} = e^{2 h(y)} \delta_{\alpha\beta}$.
In this case, the self-duality equation \eq{self-dual-eq} is exactly the same as the well-known case
on $\mathbb{R}^4$ since it is conformally invariant. Thus one may use the instanton solution \eq{inst-den}
centered at the origin of $\mathbb{R}^4$. Then the condition \eq{cinst-den} reduces to
\begin{equation}\label{weyl-h}
h(r) =  \ln \frac{\zeta \zeta_c}{r^2 + \zeta^2}
\end{equation}
where $r = \sqrt{y^\alpha y^\alpha}$.
This means that the metric on $X_4$ is given by
\begin{equation}\label{int-metric}
  ds_{X_4}^2 = h_{\alpha\beta} (y) dy^\alpha dy^\beta = e^{2 h(y)} dy^\alpha dy^\alpha =
  \frac{(\zeta \zeta_c)^2}{(r^2 + \zeta^2)^2} dy^\alpha dy^\alpha
\end{equation}
and so the internal space is precisely the four-dimensional sphere, i.e. $X_4 = \mathbb{S}^4$
with the Ricci scalar $R_{(h)} = \frac{48}{\zeta_c^2} = \zeta_c^2 \rho_1$.
Moreover, the instanton is uniformly distributed on $\mathbb{S}^4$.\footnote{It implies that
the Yang-Mills instanton on $X_4 = \mathbb{S}^4$ has no position moduli.
Thereby a nice feature of our model is to avoid the presence of any wandering instantons and achieve
a moduli stabilization except the size modulus required for the dynamical compactification
of extra dimensions.}
Therefore let us try the following metric to solve the above equations:
\begin{equation}\label{frw-metric}
    ds^2 = -dt^2 + e^{2H(t)} d\mathbf{x} \cdot d\mathbf{x} +  e^{2f(t)
    + 2 h(r)} \delta_{\alpha\beta} dy^\alpha dy^\beta,
\end{equation}
where we have assumed the FRW metric for our four-dimensional spacetime.\footnote{In our metric ansatz \eq{frw-metric},
$H(t) = \ln a(t)$ where $a(t)$ is a usual scale factor of the four-dimensional FRW universe,
so $\dot{H} = \frac{\dot{a}}{a}$ corresponds
to a Hubble parameter. Accordingly, $\ddot{H} = \frac{\ddot{a}}{a}
- \frac{\dot{a}^2}{a^2}$.}
We will show that \eq{weyl-h} is indeed a solution of the Einstein equation \eq{eins-22}.
It is interesting to see that the Yang-Mills instanton in extra dimensions uniquely fixes
the geometry of the internal space $X_4$ and leads to a compact internal space.
We will further discuss this interesting physics in section 5.

The Einstein equation (\ref{eins-22}) for the metric \eq{frw-metric} leads to the following differential equation
\begin{eqnarray} \label{pde-22}
&& \left({h'}^2 - h'' + \frac{h'}{r} \right) \frac{2 y^\alpha y^\beta}{r^2}
+ \left(2 h'' + \frac{4 h'}{r} + {h'}^2 \right) \delta_{\alpha\beta} \xx
&=& 3 e^{2f+2h} \left(\ddot{H} + 2\dot{H}^2 + \ddot{f} + 3 \dot{f} \dot{H} + 2\dot{f}^2 \right) \delta_{\alpha\beta}.
\end{eqnarray}
Hence the function $h(r)$ must satisfy the differential equation
\begin{align}
h''(r)- h'(r)\left( \frac{1}{r} + h'(r) \right) = 0.
\end{align}
As we promised, the solution is precisely Eq. \eq{weyl-h}. Then
\begin{align}
\left(2 h'' + \frac{4 h'}{r} + {h'}^2 \right) \delta_{\alpha\beta}
= - \frac{12 \zeta^2}{(r^2 + \zeta^2)^2} \delta_{\alpha\beta} = - \frac{12}{\zeta_c^2} e^{2h} \delta_{\alpha\beta}
\end{align}
and the differential equation \eq{pde-22} reduces to
\begin{eqnarray} \label{pde-23}
E_3 := \ddot{H} + 2\dot{H}^2 + \ddot{f} + 3 \dot{f} \dot{H} + 2\dot{f}^2 + \frac{4}{\zeta_c^2} e^{-2f} = 0.
\end{eqnarray}

Similarly the Einstein equation \eq{eins-11} leads to two more differential equations
\begin{align} \label{4dim-ee}
&E_1 := 2\dot{f}^2  + 4 \dot{f} \dot{H} + \dot{H}^2 + \frac{8}{\zeta_c^2} e^{-2f}
- \frac{1}{3\zeta_0^2} e^{-4f} = 0, \xx
&E_2 := 2 \ddot{H} + 3\dot{H}^2 + 4\ddot{f} + 8 \dot{f} \dot{H} + 10 \dot{f}^2
+ \frac{24}{\zeta_c^2} e^{-2f} - \frac{1}{\zeta_0^2} e^{-4f}= 0,
\end{align}
where $\zeta_0^2 \equiv \frac{g_{YM}^2 \zeta_c^4}{384 \pi G_8}$.
Since there are three equations for two unknown functions, one may think that the system would be overdetermined.
However they are not independent of each other because one can show that $\dot E_1
+ \big( 4\dot f(t) + 3\dot H(t) \big) E_1 = \dot H(t)E_2 + 4 \dot f(t) E_3$.
Indeed this is a general property of general relativity.
Note that the equation $E_1$ consists only of terms without second-order derivatives.
Therefore we will solve the evolution equations, $E_2 = 0$ and $E_3=0$, whose initial conditions
must be chosen to be consistent with the constraint equation $E_1=0$.

It may be instructive to check the consistency of Eqs. \eq{pde-23} and \eq{4dim-ee}.
Since $\nabla^\mu_{(g)} \big (R_{\mu\nu}^{(0)} - \frac{1}{2} g_{\mu\nu} R_{(g)} \big) = 0$,
the covariant derivative on the right-hand side of Eq. \eq{eins-11} with respect to $\nabla^\mu_{(g)}$
has to identically vanish. After a little algebra, one can show that this condition is reduced to
the equation
\begin{equation} \label{check-eq}
E_4 := \dot{f} \Big( \ddot{H} + \ddot{f}  + \dot{H}^2 - \dot{f} \dot{H} - \frac{4}{\zeta_c^2} e^{-2f}
+ \frac{1}{3\zeta_0^2} e^{-4f} \Big) = 0.
\end{equation}
It is easy to see that $E_4 = \dot{f}(E_3 - E_1)$, so Eq. \eq{check-eq} is indeed satisfied.
One can also deduce the equation of motion for the scale factor $f(t)$ as
\begin{align} \label{eom-f}
\ddot{f} + 3\dot{f} \dot{H} + 4\dot{f}^2 + \frac{12}{\zeta_c^2}e^{-2f}
- \frac{2}{3\zeta_0^2} e^{-4f} = 0.
\end{align}

In our model, we will see that the dynamical behavior of the internal space and four-dimensional
spacetime is opposite, as we noted below Eq. \eq{intanton-emtensor}.
That is, a generic solution shows an interesting behavior so that
the internal space is contracting if our four-dimensional spacetime is expanding and vice versa.
A similar feature was also observed in \cite{inf-exd2,dyn-comp}, but in the case without instantons.
It may be understood as a consequence of volume-preserving diffeomorphisms in general relativity.
Therefore the expansion of our four-dimensional spacetime simultaneously accompanies a
contraction of extra dimensions.
This leads to an interesting picture that a dynamical compactification of extra dimensions may be realized
through the cosmic expansion of our four-dimensional spacetime.
However, the internal space is a compact space, i.e., the four-dimensional sphere $\mathbb{S}^4$.
Since the internal space is a compact space with a smeared instanton,
the contraction cannot last forever but should end at some critical radius.
Then it is reasonable to expect a huge quantum back-reaction from the tiny internal space
and a matter production via quantum fluctuations. We will further discuss later
how the quantum back-reaction from the tiny internal space can be used to realize
a reheating mechanism and a hot big bang at early universe.

\section{Probing solution space}

In this section, we investigate the solution space spanned by the instanton spacetime in the previous section.
First we probe the solution space with flows and then analyze a perturbative solution near the boundary of an allowed parameter space.

\subsection{Solution space with flows}

As we discussed in the previous section, the three equations, $E_1=0$, $E_2=0$ and $E_3=0$, are not independent.
Actually, $E_1=0$ plays a role of the constraint on initial values,
so we have to solve another two equations, $E_2=0$ and $E_3=0$, with an initial condition
chosen to satisfy the constraint $E_1=0$.
Indeed we will consider the flow of solutions by varying the initial conditions.

First note that the system of the differential equations is controlled by two parameters,
$\zeta_c$ and $\zeta_0$. On the one hand, $\zeta_c$ is related to the radius $R$ of the internal space
by $R=\frac{\zeta_c}{2}$ since the Ricci-scalar of $\mathbb{S}^4_R$ is given by
$R_{(h)} = \frac{12}{R^2} = \frac{48}{\zeta_c^2}$. On the other hand, $\zeta_0$ is related to the four-dimensional effective cosmological constant as follows:\footnote{As we pointed out before,
the energy-momentum tensor \eq{intanton-emtensor} induced by the Yang-Mills instanton is
no longer a cosmological constant after incorporating a gravitational back-reaction.
Nevertheless $\Lambda$ in \eq{cc} sets a typical scale of
the four-dimensional effective cosmological constant.}
$$ \zeta_0^2 = \frac{g_{YM}^2 \zeta_c^4}{384 \pi G_8} =
\frac{g_{YM}^2 \mathrm{vol} (\mathbb{S}^4_R)}{64 \pi^3 G_8}
= \frac{G_4}{G_4^{eff}} \frac{1}{\Lambda} = \frac{1}{\Lambda}$$
where $\mathrm{vol} (\mathbb{S}^4_R)= \frac{8\pi^2}{3} R^4$ is the volume of $\mathbb{S}^4_R$
and the effective four-dimensional gravitational constant $G_4^{eff}
= \frac{G_8}{\mathrm{vol} (\mathbb{S}^4_R)}$ is identified with $G_4$ in Eq. \eq{cc-action}.
To find a numerical solution for the differential equations,
let us consider a scaling for the time coordinate for convenience:
\begin{align}\label{Scaling01}
t \to t/\zeta_c.
\end{align}
With this scaling, the equations are given by
\begin{align}\label{scaled_eqs1}
&e_1 := 2\dot{f}^2  + 4 \dot{f} \dot{H} + \dot{H}^2 + 8 e^{-2f}
- \frac{1}{3\tilde\zeta^2} e^{-4f} = 0, \\
\label{scaled_eqs2}
&e_2 := 2 \ddot{H} + 3\dot{H}^2 + 4\ddot{f} + 8 \dot{f} \dot{H} + 10 \dot{f}^2
+ 24 e^{-2f} - \frac{1}{\tilde\zeta^2} e^{-4f}= 0,\\
\label{scaled_eqs3}
&e_3 := \ddot{H} + 2\dot{H}^2 + \ddot{f} + 3 \dot{f} \dot{H} + 2\dot{f}^2 + 4 e^{-2f} = 0,
\end{align}
where we defined $\tilde\zeta$ as $\tilde\zeta^2 \equiv \frac{\zeta_0^2}{\zeta_c^2}
= \frac{R_{(h)}}{48 \Lambda}$.

Now we take into account the response of a solution on a specific initial condition.
Since $H$ appears only in the form of $\dot H$ in the above equations,
the only relevant initial condition is defined by $\dot H(0)$, $\dot f(0)$ and $f(0)$.
However, $f(0)$ is determined by the constraint $e_1=0$ once $\dot H(0)$ and $\dot f(0)$ are given.
Therefore we can describe the solution space by the set of
three parameters $\dot H(0)$, $\dot f(0)$ and $\tilde\zeta$.

\begin{figure}
\centering
\includegraphics[width=0.5\textwidth]{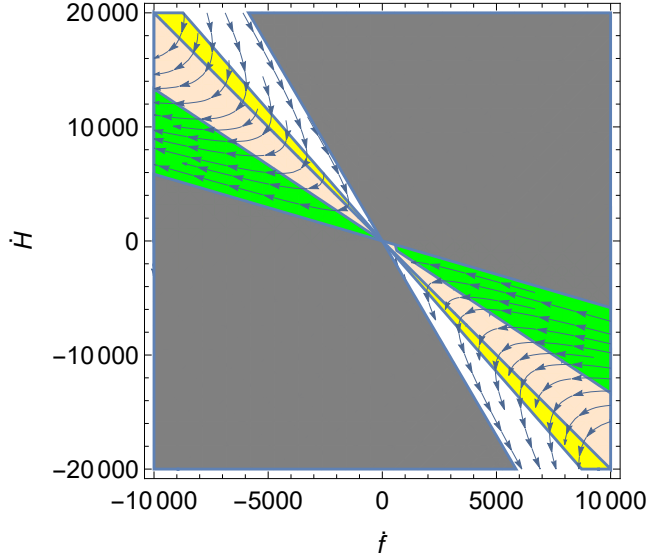}
\caption{\label{solutionflow01} The flow of solutions with $\tilde{\zeta}=\infty$:
white ($\ddot{f}>0$, $\ddot{a}<0$ , $\ddot{H}<0$)
, yellow ($\ddot{f}<0$, $\ddot{a}<0$ , $\ddot{H}<0$)
, light orange ($\ddot{f}<0$, $\ddot{a}>0$ , $\ddot{H}<0$), green ($\ddot{f}<0$, $\ddot{a}>0$ , $\ddot{H}>0$).}
\end{figure}

In order to get some intuition of the solution space, let us examine the simplest case
in the limit $\tilde\zeta = {\frac{1}{2 R \sqrt{\Lambda}}} \to \infty$.
There are two possibilities for this limit such that the background corresponds
to a very small internal space, $R \to 0$, with
fixed $\Lambda$ or a very small cosmological constant, $\Lambda \to 0$, with fixed $R$.
In this approximation, the equations are simplified as follows:
\begin{align}\label{scaled_eqs}
&\tilde e_1 := 2\dot{f}^2  + 4 \dot{f} \dot{H} + \dot{H}^2 + 8 e^{-2f}= 0, \\
&\tilde e_2 := 2 \ddot{H} + 3\dot{H}^2 + 4\ddot{f} + 8 \dot{f} \dot{H} + 10 \dot{f}^2
+ 24 e^{-2f} = 0,\\
&\tilde e_3 := \ddot{H} + 2\dot{H}^2 + \ddot{f} + 3 \dot{f} \dot{H} + 2\dot{f}^2 + 4 e^{-2f} = 0.
\end{align}
Whenever $\dot H(0)$ and $\dot f(0)$ are given, the initial information $f(0)$ is determined by $\tilde e_1=0$.
In this case, the evolution equations are given by
\begin{align}
\ddot f = - \dot{f}^2 + 3 \dot f \dot H + \frac{3}{2} \dot{H}^2,
\qquad \ddot H = - 4 \dot f \dot H - 3 \dot{H}^2.
\end{align}
Given a consistent initial condition, one can find the flow of solutions by solving the above equations.
We present the flow of solutions in Fig. \ref{solutionflow01} where $a(t) \equiv e^{H(t)}$
and $b(t) \equiv e^{f(t)}$. Once we choose a point in the space $(\dot f, \dot H)$
as an initial condition, the flow follows the curve
which passes through the point in the figure.
The gray regions are excluded by the constraint equation $2\dot{f}^2  + 4 \dot{f} \dot{H} + \dot{H}^2<0$
coming from $\tilde e_1=0$. In the lower right region, $\dot f$ is positive and $\dot H$ is negative,
so the internal space having the instanton is expanding while our (3+1)-dimensional spacetime is shrinking.
Hence this region may not be of interest.
However, the upper left region has positive $\dot H$ and negative $\dot f$.
Thus this region may be interesting since the solution space would be relevant to
a dynamical compactification of the internal space through the inflationary spacetime.

The solutions start from the white region where our universe is
in the phase of decerelating expansion ($\ddot a < 0$) while the internal space undergoes
an accelerating contraction and then its contraction is decelerating in the yellow subregion.
After that, the solutions enter to the light orange and green subregions that show an accelerating
universe ($\ddot a> 0$) with a decelerating internal space.
In the latter regions, our universe expands more rapidly while the internal space shrinks more slowly.
Therefore all the initial conditions in the upper left region result in (3+1)-dimensional inflating spacetimes
with a contracting internal space. Since Fig. \ref{solutionflow01} describes the case with a very small
internal space, it would correspond to a final stage of the inflation if it started
with a very large internal space.

In this limit, we can find a special solution when $\dot H = - \left(2  \pm  \sqrt{2}\right) \dot f$. This is nothing but the boundary in Fig. \ref{solutionflow01}, where $e^{-2f(t)}$ vanishes.
Thus the size of the internal space is infinite but decreasing.
We choose the minus sign to describe the inflating case in the green part of Fig. \ref{solutionflow01}.
In this case we find the solution as the form
\begin{align}\label{linesol01}
\dot H = \frac{v_H}{1 -\left(1+2\sqrt{2}\right)v_H t }, \qquad
\dot f = -\frac{v_H}{(2-\sqrt{2})- (3\sqrt{2}-2) v_H t}
\end{align}
where $v_H := \dot H(0)$.
Using this solution, one can compute the e-folding number $\mathcal N$ until $t=t_f$ as follows:
\begin{align}
\mathcal{N} = \int_{0}^{t_f} \dot H dt = - \frac{1}{1 + 2 \sqrt{2}}
\ln \left( 1 -(1 + 2\sqrt{2})v_H t_f \right).
\end{align}
Suppose that the initial point $(\dot f(0),\dot H(0))$ is slightly inward from the boundary.
Then one can examine the volume of the internal space and its evolution.
The e-folding number in this case is not much different from the above consideration.
This seems to imply a relation between the e-folding number and the size of internal space
at the end of inflation.

\begin{figure}
\centering
\includegraphics[width=0.45\textwidth]{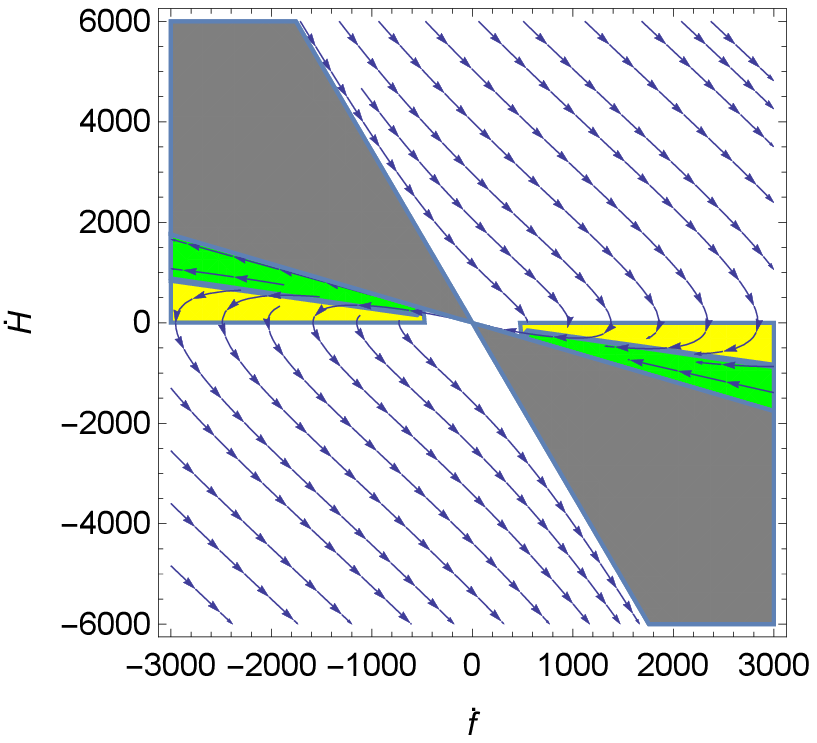}
\includegraphics[width=0.45\textwidth]{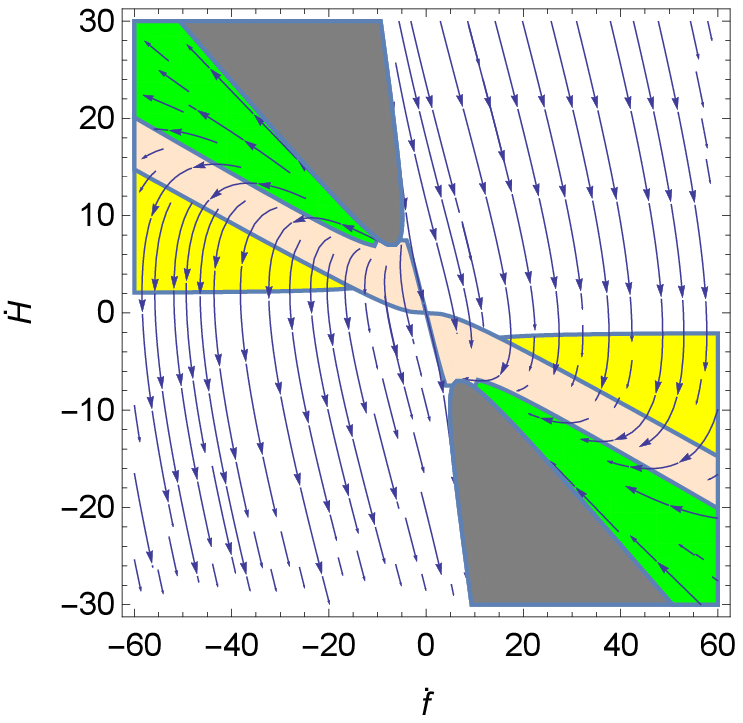}
\caption{\label{solutionflow02} The flow of solutions with $e^{-2 f}= X_-$: white ($\ddot{f}>0$, $\ddot{a}<0$, $\ddot{H}<0$), yellow ($\ddot{f}<0$, $\ddot{a}<0$, $\ddot{H}<0$),
light orange ($\ddot{f}<0$, $\ddot{a}>0$, $\ddot{H}<0$), green ($\ddot{f}<0$, $\ddot{a}>0$, $\ddot{H}>0$).}
\end{figure}

We move on to the solution space with finite $\tilde \zeta_0$.
To find a solution in this case, it is convenient to consider the following scaling symmetry
\begin{align}\label{Scaling02}
t \to \tilde\zeta t~~, \qquad e^{-2f}\to \frac{1}{\tilde\zeta^2} e^{-2f}
\end{align}
for the set of the equations, $\{e_1,e_2,e_3\}=0$.
Then the differential equations take the following forms
\begin{align}
&\bar e_1 := 2\dot{f}^2  + 4 \dot{f} \dot{H} + \dot{H}^2 + 8 e^{-2f}
- \frac{1}{3} e^{-4f} = 0, \\
&\bar e_2 := 2 \ddot{H} + 3\dot{H}^2 + 4\ddot{f} + 8 \dot{f} \dot{H} + 10 \dot{f}^2
+ 24 e^{-2f} - e^{-4f}= 0,\\
&\bar e_3 := \ddot{H} + 2\dot{H}^2 + \ddot{f} + 3 \dot{f} \dot{H} + 2\dot{f}^2 + 4 e^{-2f} = 0,
\end{align}
which are equal to the previous Eqs. \eq{scaled_eqs1}-\eq{scaled_eqs3} with $\tilde\zeta=1$.

In order to find solutions, we need to impose initial conditions obeying the constraint $\bar e_1=0$.
Once a point in the space $(\dot f, \dot H)$ is chosen as an initial condition,
as we did in the previous analysis, one can then solve the equation, $\bar e_1=0$,
to find an initial value $f(0)$:
\begin{align}
e^{-2f}=12 \mp \sqrt{3} \sqrt{4 \dot{f} \dot{H} + 2 \dot{f}^2+\dot{H}^2+48} \equiv X_\pm.
\end{align}
By plugging these into $\bar e_2=0$ and $\bar e_3=0$, one can construct the flow equations
for $(\ddot f, \ddot H)$. First let us consider the case with $e^{-2f}=X_-$.
The flow equations are given by
\begin{align}
&\ddot f = 2 \dot{H}^2 + 5 \dot{f} \dot{H}+4 \sqrt{3} \sqrt{4 \dot{f} \dot{H}+2 \dot{f}^2
+ \dot{H}^2+48} + 48, \label{solutionflowXm}\\
&\ddot H =-2 \dot{f}^2-4 \dot{H}^2 -8 \dot{f} \dot{H}-8 \sqrt{3} \sqrt{4 \dot{f} \dot{H}
+ 2 \dot{f}^2+\dot{H}^2+48} -96.\nonumber
\end{align}
The flow of solutions is shown up in Fig. \ref{solutionflow02}.
In this figure the gray subregions are excluded again since $e^{-2f}$ becomes a complex number there.
One can see that all the solutions flow into the region with $\dot f >0$ and $\dot H<0$.
For this reason, we are not interested in this case.

\begin{figure}
\centering
\includegraphics[width=0.58\textwidth]{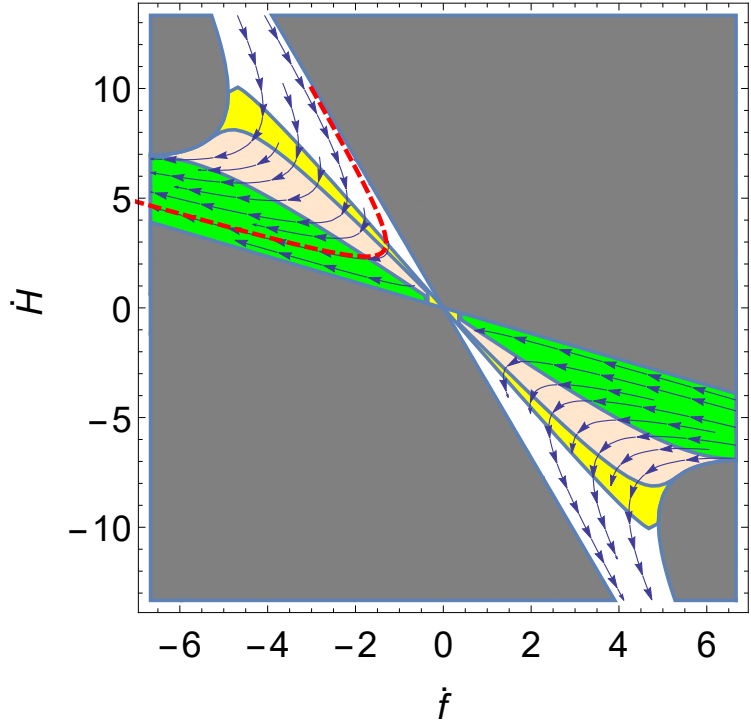}
\caption{\label{solutionflow03} The flow of solutions with $e^{-2 f}= X_+$:
The red dashed line depicts a solution curve shown in Fig. \ref{fH_m3_10}.
Each subregion denotes:
white ($\ddot{f}>0$, $\ddot{a}<0$, $\ddot{H}<0$),
yellow ($\ddot{f}<0$, $\ddot{a}<0$, $\ddot{H}<0$),
light orange ($\ddot{f}<0$, $\ddot{a}>0$, $\ddot{H}<0$),
and green ($\ddot{f}<0$, $\ddot{a}>0$, $\ddot{H}>0$).}
\end{figure}

The other situation is the case with $e^{-2f}=X_+$.
Here, one may notice an interesting feature that the size of internal space has the minimum value given by $e^{2f}=1/12$,
i.e. $e^{f} \geq \frac{1}{2 \sqrt{3}}$.
Similarly one can get the flow equations for $( \ddot f, \ddot H )$ given by
\begin{align}
&\ddot f = 5 \dot{f} \dot{H}+2 \dot{H}^2+48 -4 \sqrt{3} \sqrt{4 \dot{f} \dot{H}+2 \dot{f}^2+\dot{H}^2+48}, \label{solutionflowXp}\\
&\ddot H = -96-2 \left(4 \dot{f} \dot{H}+\dot{f}^2+2 \dot{H}^2\right)+8 \sqrt{3} \sqrt{4 \dot{f} \dot{H}+2 \dot{f}^2+\dot{H}^2+48}. \nonumber
\end{align}
The result in this case is shown up in Fig. \ref{solutionflow03}.
Some excluded regions also appear. In the upper left and lower right regions, $e^{-2f}$ becomes a complex number so that they are not physically allowed. In addition, in the lower left and upper right regions, $e^{-2f}$ becomes negative, so they are not allowed either as a physical solution space.
Solutions with consistent initial conditions $\dot H >0$ and $\dot f <0$ flow to the solution space
of $\ddot a >0$ and $\ddot f <0$ finally.
Therefore, we can obtain (3+1)-dimensional inflating spacetimes with a contracting internal space.
For this reason, the upper left side is a welcomed place for our universe.

\begin{figure}
\centering
\includegraphics[width=0.45\textwidth]{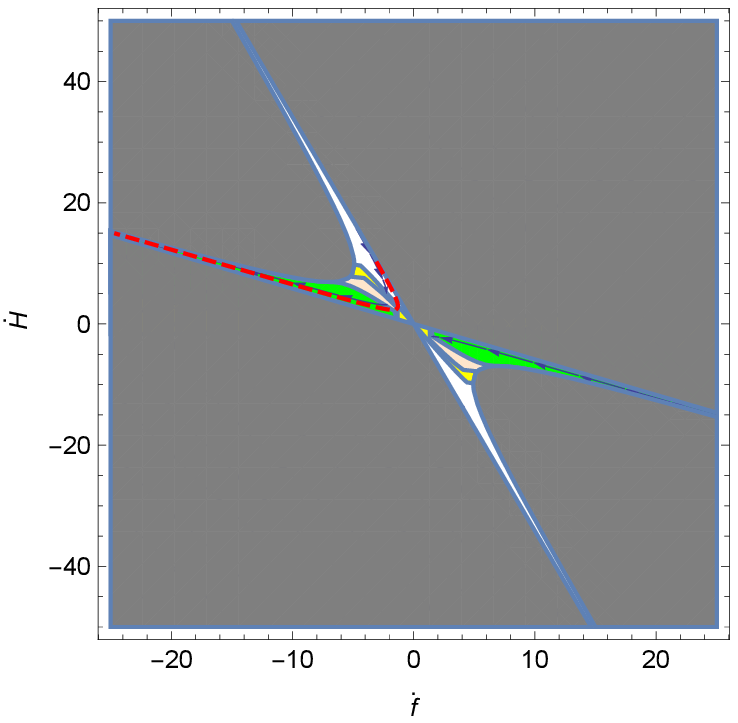}
\includegraphics[width=0.45\textwidth]{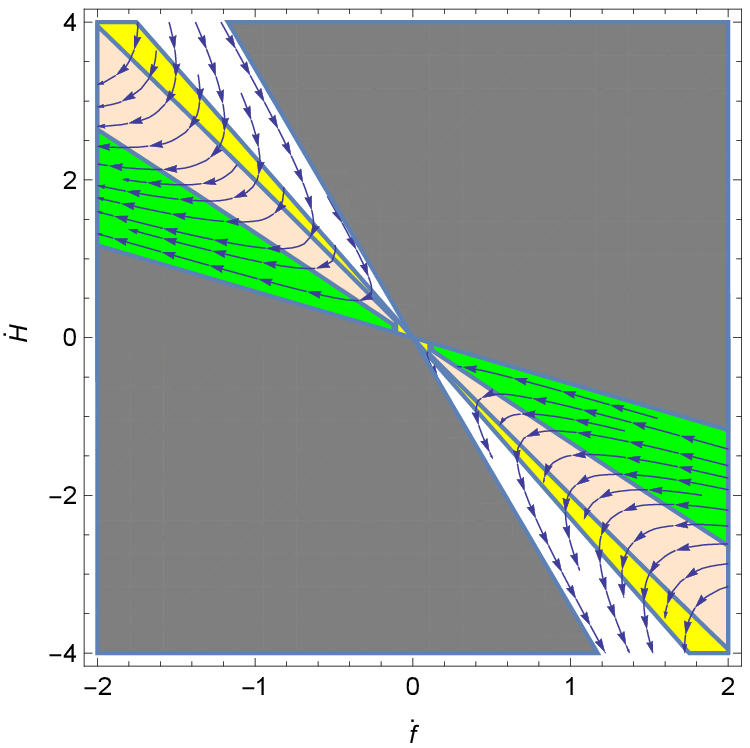}
\caption{\label{solutionflow03-12} The flow of solutions with $e^{-2 f}= X_+$: The red dashed line represents a solution curve shown in Fig. \ref{fH_m3_10}. This is the same flow solution in Fig. \ref{solutionflow03} except the size of the parameter space $(\dot f, \dot H)$.}
\end{figure}

\begin{figure}
\centering
\includegraphics[width=0.4\textwidth]{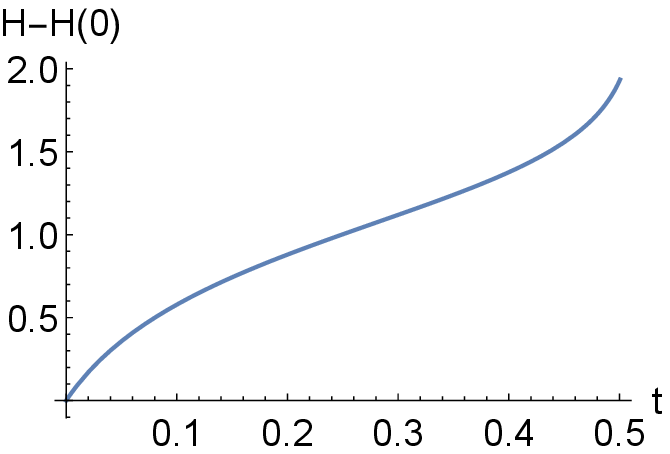}
\includegraphics[width=0.4\textwidth]{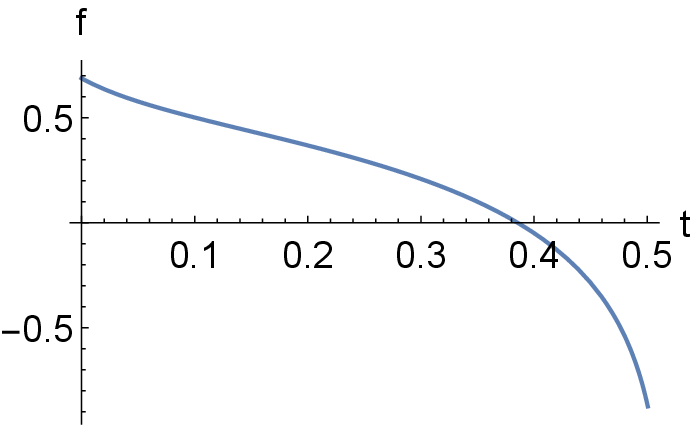}
\caption{\label{fH_m3_10}
$H$ and $f$ for the red solid curve in Fig. \ref{solutionflow03} and the red dashed curve in the left figure of Fig.\ref{solutionflow03-12} with $\dot f(0)= -3$ and $\dot H(0)=10.$ }
\end{figure}

We would like to briefly comment on the scaling property of solutions.
Since we have scaled the time coordinate by
\begin{align}\label{timeS}
t \to t' = \frac{\tilde\zeta}{\zeta_c} t,
\end{align}
considering a narrow parameter space of $(\dot f, \dot H)$ corresponds to considering the solution space with a larger $\tilde \zeta$. In Fig. \ref{solutionflow03-12}, we display the flow of solutions with a wider parameter region and
a narrower parameter region, respectively. One can notice that the figure for the narrow parameter region is very similar to Fig. \ref{solutionflow01} for $\tilde{\zeta} = \infty$ case. The scaled time $t'=\sqrt{\frac{g_{YM}^2}{384\pi G_8}} t$
is dimensionless and depends only on the ratio of the eight-dimensional couplings in Yang-Mills theory and gravity.
In addition, solutions depend on two more parameters, $H(0)$ and $\zeta_c = 2R$, which govern
the initial sizes of the (3+1)-dimensional spacetime and the four-dimensional internal space.
It is reasonable to imagine that the volumes of two spaces are comparable each other at the beginning.
A typical solution for such a background is shown in Fig. \ref{fH_m3_10}, whose solution curve is indicated
by the dashed line in Fig. \ref{solutionflow03} and the left figure in Fig. \ref{solutionflow03-12}.

Now, let us discuss the late time behavior of solutions.
As Fig. \ref{solutionflow03-12} indicates, the allowed region of the solution space gets narrower
as $\dot H$ becomes lager.
One can easily show that the width of the allowed region for negative $\dot f$ is given
by $\Delta \dot H = \sqrt{2}\left( |\dot f| - \sqrt{\dot{f}^2-24} \right)$.
The upper boundary of the allowed region is determined by the equation, $4 \dot{f} \dot{H}+2 \dot{f}^2+\dot{H}^2+48=0$,
which also appears in the flow equation (\ref{solutionflowXp}).
For a large $\dot H > 0$, the boundary is given by $\dot H \sim -\left(2 \pm\sqrt{2}\right) \dot{f}$.
Both branches correspond to solutions with a (3+1)-dimensional expanding spacetime $(\dot H > 0)$ but a contracting
four-dimensional internal space  $(\dot f < 0)$. In this case, the flow equation becomes
\begin{align}
(\ddot f\,,\,\ddot H) \sim   \left((2 \pm \sqrt{2}) \dot{f}^2, \; (-10 \mp 8 \sqrt{2}) \dot{f}^2 \right).
\end{align}
It leads to the behavior $\ddot a(t)\sim  4(-1 \mp \sqrt{2})  e^{H(t)} \dot f(t)^2$
and $\ddot b(t)\sim  3( 1 \pm \sqrt{2})  e^{f(t)} \dot f(t)^2$.
Therefore, in the minus-branch, the expansion of the (3+1)-dimensional spacetime
is accelerating whereas the contraction of the four-dimensional internal space is decelerating
but, in the plus-branch, the behavior is opposite.
However, the contraction of the four-dimensional internal space in the minus-branch cannot be forever
because the internal space is a compact space, i.e., the four-dimensional sphere $\mathbb{S}^4$,
where the instanton is supported.
It is reasonable to expect that quantum gravity effect cannot be ignored as the internal space
is getting smaller and smaller.
If so, a huge quantum back-reaction from the tiny internal space will arise at some critical radius.
Due to the back-reaction, the internal space may be oscillating around the critical radius.
As a result, some matters may be produced by quantum fluctuations.
Thus it is necessary to turn on the fluctuations of Yang-Mills gauge fields around
the instanton background \eq{8-gauge} as well as metric fluctuations
near the background metric \eq{frw-metric}.
It will be interesting to consider the effect of quantum fluctuations coupled
to the Einstein-Yang-Mills theory with an instanton background.

\subsection{Perturbative solution}

We found an interesting behavior of
generic solutions showing that the internal space is contracting if our four-dimensional spacetime
is expanding and vice versa. Since the internal space in our case is not an empty space but supports
the Yang-Mills instanton contrary to the situation in \cite{inf-exd2,dyn-comp},
there may be naturally a large quantum back-reaction (a.k.a. reheating)
after the dynamical compactification of
extra dimensions via the inflation in higher-dimensional spacetimes such as string theory.
In this subsection, we focus on the case with $e^{-2f}=X_+$ and find an approximate solution.
Although our model is a preliminary step to apply to the ten-dimensional string theory,
this solution may give us a more clear way to constrain the parameters of the model from the cosmological data.

In the previous subsection, we found that there is an exact solution flow (\ref{linesol01}) moving along the boundary of the solution space in the limit $\tilde\zeta \to \infty$. In this limit, it is easy to show that Eq. (\ref{solutionflowXp}) satisfying the relation $\dot H = - (2-\sqrt{2})\dot f$ reduces to the equations
\begin{align}
\ddot f = (2-3\sqrt{2})\dot f^2, \qquad \ddot H =(1+2\sqrt{2})\dot H^2,
\end{align}
These equations are not independent of each other since they are related by the constraint $\dot H = - (2-\sqrt{2})\dot f$.
The solution is given by (\ref{linesol01}) and this solution may be extended to a finite but large $\tilde \zeta$.
So we would like to find another solution that is slightly off the lower boundary. To get such a solution, we introduce a following perturbation
\begin{align}
\dot H(t) = \dot H_b(t) + \lambda\, \delta
\dot H(t),  \qquad \dot f(t) = \dot f_b(t) + \lambda\, \delta
\dot f(t),
\end{align}
where $\dot H_b(t)$ and $\dot f_b(t)$ denote the lower boundary solution (\ref{linesol01}) and $\lambda$ is a small expansion parameter. We will focus on the linear order in $\lambda$ from now on.
Then the linearized equations are given by
\begin{align}
\delta  \ddot{f}=\left(\sqrt{2}+5\right)  \dot{H}_b\,\delta\dot{f}-\frac{3  \dot{H}_b}{\sqrt{2}}\delta\dot{H}, \qquad \delta  \ddot{H}=2 \dot{H}_b \left((\sqrt{2}-1) \delta  \dot{H}-2 \delta  \dot{f}\right).
\end{align}
The above equations are a coupled linear differential equation. The solution can be obtained as follows:
\begin{align} \label{linpert-f}
&\delta \dot{f}= \frac{\sqrt{2} v_H^2 \left(3 C_2 \left(1-\left(2 \sqrt{2}+1\right)v_H t \right)^{\frac{1}{7} \left(19-3 \sqrt{2}\right)}+C_1\right)}{\left(1-2 \sqrt{2}\right)^2 \left(1-\left(2 \sqrt{2}+1\right) v_H t\right)^2}, \\
 \label{linpert-H}
&\delta \dot{H} = \frac{2 v_H^2 \left( 4 \left( 3 + 2 \sqrt{2} \right)  C_2 \left(1-\left(2 \sqrt{2}+1\right) v_H t \right)^{\frac{1}{7} \left(19-3 \sqrt{2}\right)}-\sqrt{2} C_1\right)}{\left(\sqrt{2}+10\right) \left(1-\left(2 \sqrt{2}+1\right) v_H t\right)^2},
\end{align}
where the integration constants are given in terms of the initial condition:
\begin{align}
C_1 = \frac{4\, \delta  \dot{f}(0) + 3 \left(\sqrt{2}-1\right) \delta  \dot{H}(0)}{2  v_H^2}, \qquad C_2=\frac{\left(3 \sqrt{2}-4\right) \delta  \dot{f}(0)+\left(\sqrt{2}-1\right) \delta  \dot{H}(0)}{2 v_H^2}.
\end{align}

Now one may examine an early stage of inflation. The inflation is defined by the condition $\ddot a > 0$ where $a(t)=e^{H(t)}$.
This condition can be written as
\begin{align}
e^{-H}\ddot a = \left(\ddot H +\dot{H}^2 \right) > 0.
\end{align}
From this equation one can determine a constraint of initial conditions for the beginning of inflation. However the linear approximation is not enough to describe the evolution near the beginning. Thus we have to wait some time until the linear approximation becomes valid. During this period, some amounts of e-folding would be accumulated. We denote this e-folding by $\mathcal{N}_i$. For a convenience, we may set $t=0$ when the linear approximation starts to be valid. Then the total e-folding can be written as follows \cite{kolb-turner,Baumann:2014nda}
\begin{align}
\mathcal{N} =\mathcal{N}_i + \int_0^{t_f} dt \dot H = \mathcal{N}_i + \int_0^{t_f} dt \dot{H}_b + \lambda \int_0^{t_f} dt \,\delta\dot H + \mathcal{O}\left(\lambda^2\right).
\end{align}
The above integration can be evaluated to determine the e-folding number $\mathcal{N}$ and it is given by
\begin{align}
\mathcal N \sim \mathcal N_i + \mathcal I_0 + \lambda\, \mathcal I_1,
\end{align}
where
\begin{align}
\mathcal I_0 =&-\frac{\ln \left(1-\left(2 \sqrt{2}+1\right) v_H t_f\right)}{2 \sqrt{2}+1},\\
\mathcal I_1 =&-\frac{2 \left(4 \sqrt{2}+9\right) \left(1-\left(2 \sqrt{2}+1\right) v_H t_f\right){}^{\frac{3}{7} \left(4-\sqrt{2}\right)}}{147 v_H}\nonumber\\& + \frac{3 \left(4 \sqrt{2} - 5 \right)}{49 v_H \left(1-\left(2 \sqrt{2}+1\right) v_H t_f\right)}+\frac{41 \sqrt{2}-31}{21 \left(5 \sqrt{2}+1\right) v_H}.
\end{align}
The figure \ref{solutionflow03} indicates that the accelerating expansion of our four-dimensional
spacetime may be lasted for a sufficiently long period although we do not have enough information
about the rapidity of the flow and the status of the last stage of the inflation.
If the expansion is lasted for until $t_f \approx \frac{0.26}{v_H}$,
the e-folding number could be sufficiently large, possibly, $\mathcal N \sim 60$.

\begin{figure}
 \centering
\includegraphics[width=0.42\textwidth]{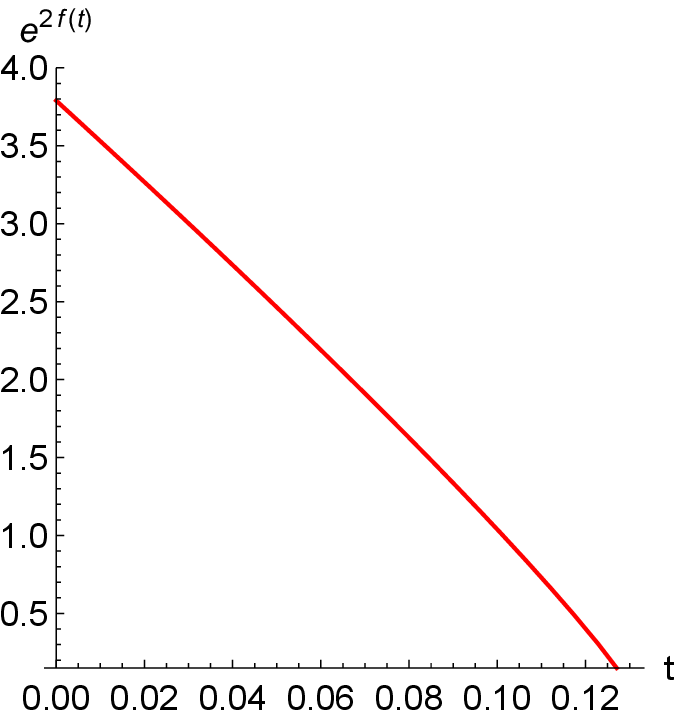}
 \includegraphics[width=0.49\textwidth]{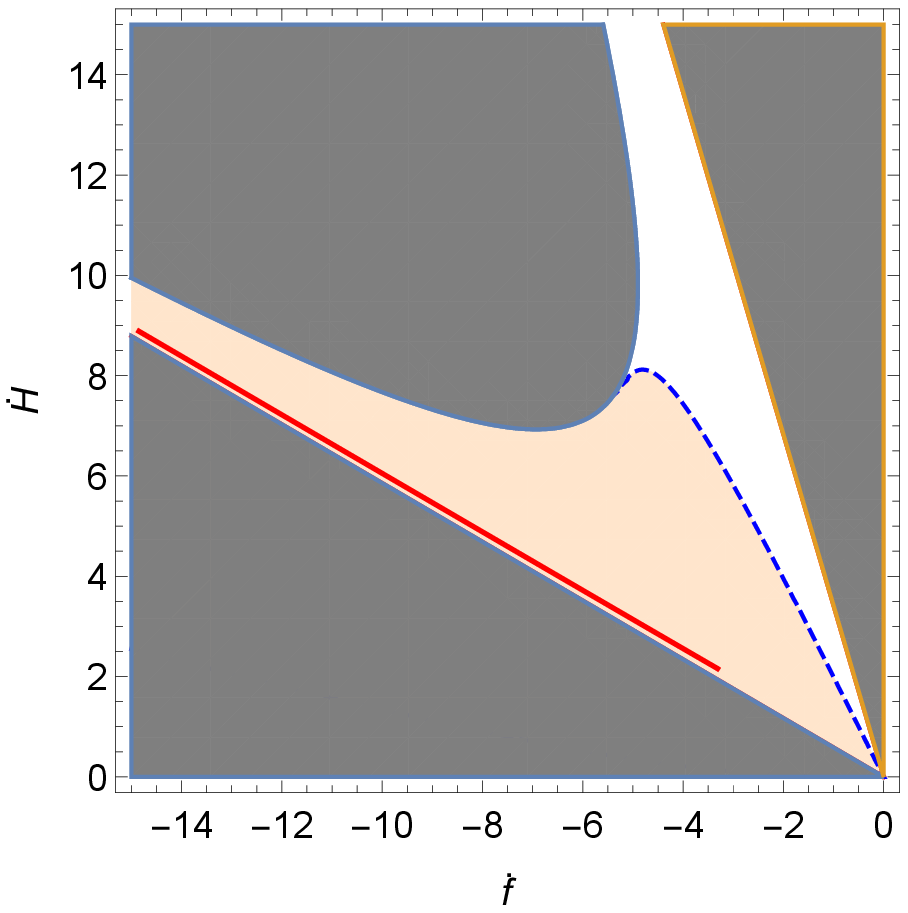}
 \caption{\label{2p2f}
A scale factor $e^{2f(t)}$ (left) and corresponding solution trajectory (right) in the linear approximation: The initial conditions are chosen as $\dot{H}(0)=2$, $\delta\dot{f}(0)= 1$ and $\delta\dot{H}(0)=1.6$ with $\lambda= 10^{-1}$ for visualization. The light orange region stands for $\ddot a>0$
and thus the dashed blue line can be regarded as a set of starting initial points for inflation.}\label{ep2f}
 \end{figure}

Another important physical quantity is the size of internal space. The length scale corresponding to the space is given by $e^{f(t)}\zeta_c /\tilde\zeta$ which is determined by solving the constraint $e^{-2f}=X_+$.
Using the linear solution \eq{linpert-f}, one can compute the leading order of $e^{2f}$ to yield
\begin{align}
e^{2f (t)} \approx  \frac{4(2-\sqrt{2})}{\lambda \Big(2 (\sqrt{2}-1)\delta  \dot{f} + \sqrt{2} \delta  \dot{H} \Big) \dot H_b}.
\end{align}
We plot this function in Fig. \ref{ep2f} with a specific initial condition. One can see that the internal space size is rapidly decreasing but it could not be smaller than $\frac{\zeta_c}{2 \sqrt{3}\tilde\zeta}
= \sqrt{\frac{32 \pi G_8}{g^2_{YM}}}$ because $e^{f} \geq \frac{1}{2 \sqrt{3}}$.
Quantum gravity effects may not be ignored when the scale $ e^{f (t)} \zeta_c /\tilde\zeta$ is comparable with the Planck length.
According to a simple dimensional analysis, we have the relation $G_8 = g^2_{YM} l_s^2$
with a typical microscopic scale $l_s$ where the quantum gravity effect becomes significant.
Therefore one can expect that there will be a large quantum back-reaction from the tiny internal space
with the size of roughly $\sim 10 l_s$. Then there will be an oscillating period
as a result of the gravitational back-reaction,
and the inflation would stop with a large amount of matter production
via the coupling of the oscillating modes with Yang-Mills gauge fields and metric fluctuations in Einstein-Yang-Mills theory. This phenomenon would correspond to a reheating mechanism
in inflationary cosmology. Thus we speculate that the internal space can be stabilized through
the quantum back-reaction from the tiny internal space and
the stabilized length scale may be the same order of
$e^{f(t_f)}\zeta_c/\tilde\zeta \approx \frac{\zeta_c}{2 \sqrt{3}\tilde\zeta}
= \sqrt{\frac{32 \pi G_8}{g^2_{YM}}} \sim 10 l_s $. It means that the dynamical relaxation of the internal
space is closely correlated with the matter generation in our four-dimensional spacetime
via the reheating mechanism.
If our speculation is true, our toy model simultaneously provides the dynamical compactification of
extra dimensions through the cosmic inflation of our four-dimensional spacetime as well as
the reheating mechanism in inflationary cosmology via the quantum back-reaction
from the tiny internal space. Then our universe will finally enter into a hot big bang
in radiation-dominated era with the internal space of Planck size.

When the internal space experiences a large quantum back-reaction,
it enters to a phase of UV physics with a strong gravity since quantum fluctuations
would be blue-shifted due to the contraction of  the internal space.
However it is well-known \cite{kolb-turner,Baumann:2014nda}
that quantum fluctuations in four-dimensional spacetime would
be red-shifted during the inflation
and generate an almost scale-invariant spectrum of density perturbations.
Therefore the inflating spacetime would be dominated by the IR physics.
This implies that the UV physics in internal space may be deeply related to the IR physics
in four-dimensional spacetime at late times of inflation.
Note that non-Abelian gauge theory in four dimensions is weakly coupled in UV regime
while four-dimensional gravity is strongly coupled. However, this behavior is reversed at IR.
It further implies that the role of gravity and gauge theory in the internal space
and four-dimensional spacetime would be very different for the reheating mechanism.
This feature is very reminiscent of the UV/IR interplay discussed in several literatures \cite{uv-ir}.

\section{Discussion}

We have considered Yang-Mills instantons on a four-dimensional internal space $X_4$ whose
instanton number is given by
\begin{eqnarray}\label{2nd-chern}
    k &=& - \frac{1}{16 \pi^2} \int_{X_4} \mathrm{Tr} F \wedge F \in \mathbb{Z} \xx
    &=& \mp \frac{1}{32 \pi^2} \int_{X_4} d^4 y \sqrt{h}  h^{\alpha\gamma} h^{\beta\delta}
    \mathrm{Tr} F_{\alpha\beta} F_{\gamma\delta}.
\end{eqnarray}
In this case, the instanton action $S=\frac{8\pi^2 |k|}{g_{YM}^2}$ is metric-independent
as Eq. \eq{2nd-chern} indicates \cite{rajaraman}.
We have observed in section 3 that the dynamical spacetime generated by Yang-Mills instantons in $X_4$
is consistent only if the instanton density is constant, i.e., $\rho_n (y)=$ constant;
in our case, $ \rho_1 (y) = 48/\zeta_c^4$.
This condition has uniquely fixed the internal space $X_4$ as
the four-dimensional sphere $\mathbb{S}_R^4$ with radius $R=\frac{\zeta_c}{2}$.
In this case, the instanton action is simply proportional to the volume of the internal
space $X_4$. To be specific, it is given by
\begin{equation}\label{inst-vol}
S= \frac{1}{g_{YM}^2} \int_{X_4} d^4 y \sqrt{h} \rho_n (y)= \frac{\rho_n \mathrm{vol} (X_4)}{g_{YM}^2}.
\end{equation}
For the solution \eq{weyl-h}, for example, the volume of $X_4$ is given by
$\mathrm{vol} (X_4) = \int_{X_4} d^4 y \sqrt{h} = \frac{\pi^2 \zeta_c^4}{6}$.
 However, this result suggests several interesting implications
and leads to questions.

Suppose that there was no instanton at the beginning, so $T_{\mu\nu} = 0$
in Eq. \eq{geom-x}. Then a natural vacuum geometry is the eight-dimensional
flat Minkowski spacetime $\mathbb{R}^{7,1}$ although a general vacuum geometry is
of the form $\mathcal{M}_8 = \mathbb{R}^{3,1} \times X_4$ with a Ricci-flat manifold $X_4$
such as K3 or any hyper-K\"ahler manifolds. Assume that, after a while,
Yang-Mills instantons of the type \eq{8-gauge} are formed in the internal space $X_4$.
We have analyzed what happens in the eight-dimensional spacetime
after the instanton formation. As a result of the instanton formation,
the internal space $X_4$ is compactified from $\mathbb{R}^4$ to $\mathbb{S}^4$.\footnote{Although the vacuum transition
from no instanton to instantons seems to be an energy-violating process at first sight,
we have checked at the end of section 3 that this kind of vacuum transition becomes consistent
after incorporating the gravitational back-reaction.
In our case, the existence of instantons is an initial condition
at an early universe as a consistent vacuum solution of the eight-dimensional
Einstein-Yang-Mills theory \eq{total-action}.
We did not address the issue on the vacuum transition.
We need a quantum gravity theory to understand
how the vacuum transition is physically realized as a nonperturbative time evolution.}
Furthermore, there is a branch where the size of the internal space is decreasing
when our four-dimensional spacetime is expanding.
Note that this compactification is a topology-changing process.
But there is a well-known theorem \cite{sing-theorem} showing that generic topology-changing
spacetimes are singular. Thus such a topology change does not seem to be allowed in classical
general relativity and it necessarily accompanies a quantum gravity
effect beyond the general relativity. Hence this topology-changing process may be possible only at early universe where quantum gravity effect is very strong.

Our original motivation of instanton induced inflation was to overcome the difficulties of stringy inflation using scalar fields, i.e. moduli fields,
that we have discussed in section 1. Our toy model may be embedded into string theory by introducing a two-dimensional torus $\mathbb{T}^2$ as a spectator. So the ten-dimensional target spacetime becomes $\mathcal{M}_8 \times \mathbb{T}^2$. We need to embed the four-dimensional instanton solution \eq{8-gauge}
into a ten-dimensional Yang-Mills gauge theory. Actually it can be done by the simple ansatz
\begin{equation}\label{10-gauge}
    A_\mu (x, y, z) = 0, \qquad A_\alpha (x,y, z) = A_\alpha (y), \qquad  A_i (x, y, z) = 0,
\end{equation}
where $z^i \; (i=1,2)$ are coordinates on $\mathbb{T}^2$. The ten-dimensional Yang-Mills gauge theory may be simply embedded into heterotic string theory by considering the Yang-Mills gauge group $G$ as a subgroup of
$E_8 \times E_8$ or $SO(32)$. More interesting model which can be embedded into string theory is to consider
six-dimensional Hermitian Yang-Mills instantons \cite{hym-inst} in the ten-dimensional Einstein-Yang-Mills theory
with a gauge group $G$. The gauge group $G$ can be simply embedded into $E_8 \times E_8$ or $SO(32)$ in
heterotic string theory. One may take the ansatz
\begin{equation}\label{46-gauge}
    A_\mu (x, y) = 0, \qquad A_\alpha (x,y) = A_\alpha (y),
\end{equation}
where $y^\alpha \; (\alpha=1,\cdots, 6)$ are now coordinates on a six-dimensional manifold $X_6$, e.g., a Calabi-Yau manifold.
In this case, the ten-dimensional Yang-Mills gauge theory is reduced to the six-dimensional
gauge theory on $X_6$ whose action is similar to Eq. \eq{8-4-action}.
The Hermitian Yang-Mills instanton is defined by rewriting the six-dimensional
Yang-Mills gauge theory as the Bogomol'nyi form \cite{Yang:2011ts}
\begin{eqnarray}\label{six-ym}
  S_{YM} &=& - \frac{1}{4g_{YM}^2} \int_{X_6} d^6 y \sqrt{h} h^{\alpha\gamma} h^{\beta\delta}
\mathrm{Tr} F_{\alpha\beta} F_{\gamma\delta} \\
  &=& - \frac{1}{8 g_{YM}^2} \int_{X_6} d^6 y \sqrt{h} \mathrm{Tr}
  \left[ \Big(F_{ab} \pm * (F \wedge \Omega)_{ab} \Big)^2 - \frac{1}{2} \Big( \Omega_{ab} F^{ab} \Big)^2 \right] \pm \frac{1}{g_{YM}^2} \int_{X_6} \mathrm{Tr} F \wedge F \wedge \Omega, \nonumber
\end{eqnarray}
where $\Omega$ is a K\"ahler form on $X_6$. The action is minimized by the configuration satisfying the Hermitian Yang-Mills equation
\begin{equation}\label{hym-eq}
 F_{ab} \pm * (F \wedge \Omega)_{ab} = 0, \qquad (a,b=1, \cdots, 6).
\end{equation}
Then the stability condition is automatically satisfied, i.e. $\Omega_{ab} F^{ab}=0$,
so the action for the Hermitian Yang-Mills instantons is given by
\begin{equation}\label{hym-action}
  S_{YM} = \pm \frac{1}{g_{YM}^2} \int_{X_6} \mathrm{Tr} F \wedge F \wedge \Omega \geq 0,
\end{equation}
which is known as a topological number \cite{hym-q}. Therefore the Hermitian Yang-Mills instantons formed
in the six-dimensional internal space also generate a (quantized) cosmological constant.
Thus it is necessary to incorporate the gravitational back-reaction
from the Hermitian Yang-Mills instantons as we examined for the case of four-dimensional Yang-Mills
instantons. One crucial difference compared to the four-dimensional case is that
the energy-momentum tensor $T_{\alpha\beta}$ in Eq. \eq{emtensor-1} no longer vanishes.
This fact may allow us a localized Hermitian Yang-Mills instanton in $X_6$ as an anti-gravitating
matter source in our four-dimensional spacetime.
We expect that this model can be embedded into heterotic string theory.
We hope to report our progress along this direction in the near future.

Our model can be supersymmetrized and described by a ten-dimensional ${\cal N}=1$
heterotic supergravity \cite{het-sugra}. In this supersymmetric setup, four-dimensional
Yang-Mills instantons \eq{10-gauge} and six-dimensional Hermitian Yang-Mills
instantons \eq{46-gauge} are BPS states. Then our model may provide an interesting mechanism
for the dynamical supersymmetry breaking. As we observed in section 3,
Yang-Mills instantons in extra dimensions develop a time-dependence in the gravitational metric
after incorporating their gravitational back-reaction. The time-dependent dynamical spacetime
will break the underlying supersymmetry preserved by the BPS state.
It will be interesting to investigate in some detail the dynamical supersymmetry breaking by Yang-Mills
instantons acting as an anti-gravitating matter source.

Although we have shown that the Yang-Mills instantons in extra dimensions can trigger
the expansion of our Universe in four-dimensional spacetime
as well as the dynamical compactification of extra dimensions,
it is not clear yet whether the instanton based inflation really provides a
viable inflationary model. It would be necessary to test
the cosmological application and consistency such as
the slow roll inflation with a scale-invariant spectrum and density perturbations.
We will report our result in the follow-ups.
\\

{\bf Note added:}  After submitting this paper to arXiv, we were informed closely related works \cite{Kihara:2009ea,cz} by
Muneto Nitta and Eoin \'O Colg\'ain whom we thank for the information.
Let us comment about some crucial difference of our work from Ref. \cite{cz}.
In our paper, we have not introduced a bare cosmological constant unlike the case in \cite{cz}.
Nevertheless we have found that the internal geometry becomes $\mathbb{S}^4$ (a Euclidean de Sitter space)
due to the presence of an instanton and it is uniquely fixed by the equations of motion whereas
it was assumed in \cite{cz} from the beginning. One of the interesting features in our paper is that the dynamical compactification
of extra dimensions simultanously happens with the cosmic expansion of our four-dimensional spacetime.
We understand this behavior as a consequence of volume-preserving diffeomorphisms in general relativity.
However this conclusion is not valid with a cosmological constant which allows even a static solution $\mathbb{R}^{3,1}
\times \mathbb{S}^4$ as shown in \cite{sim-idea}.
Moreover, the dimensional reduction was performed in \cite{cz} to analyze the four-dimensional effective theory of inflation,
so the response of internal space could not be addressed.


\section*{Acknowledgments}

We acknowledge the hospitality at APCTP where part of this work was done.
This work was supported by the National Research Foundation of Korea (NRF)
with grant numbers NRF-2015R1D1A1A01058220 (K.K.)
and NRF-2015R1D1A1A01059710 (H.Y.).
S.K was supported by Basic Science Research Program through the National Research Foundation of Korea
funded by the Ministry of Education (No. NRF-2016R1D1A1B04932574).

\end{document}